\documentclass[10pt,conference, urlbreakonhyphens]{IEEEtran}

\usepackage[utf8]{inputenc}
\usepackage{hyperref}
\usepackage{textcomp}
\usepackage{makecell}
\usepackage{tabularx}
\usepackage{comment}
\usepackage{float}
\usepackage{verbatim}
\usepackage{amsmath}
\usepackage{graphicx}

\usepackage{amssymb}
\graphicspath{{./figs/}}
\usepackage[font=small,labelfont=bf]{caption}
\usepackage[rightcaption]{sidecap}
\usepackage{wrapfig}
\usepackage{subcaption}
\usepackage{multirow}
\usepackage{flexisym}
\usepackage{enumerate}
\usepackage{diagbox}
\usepackage{xspace}
\usepackage{booktabs}
\usepackage{url}
\usepackage{tcolorbox}
\usepackage{etoolbox}
\AtBeginEnvironment{tcolorbox}{\tiny}
\usepackage{xcolor,colortbl}
\usepackage{slantsc}
\usepackage{authblk}

\newcommand{\sorcerer}{\rmfamily{\scshape{So\{u\}rcerer}}}

\AtBeginDocument{%
  \providecommand\BibTeX{{%
    \normalfont B\kern-0.5em{\scshape i\kern-0.25em b}\kern-0.8em\TeX}}}

\begin{document}

\title{\sorcerer~: Developer-Driven Security Testing Framework for Android Apps}

\author{Muhammad Sajidur Rahman}
\author{Blas Kojusner}
\author{Ryon Kennedy}
\author{Prerit Pathak}
\author{Lin Qi}
\author{Byron Williams}
\affil{University of Florida, \\ Florida Institute for Cybersecurity Research, FL, USA. \authorcr Email: {\tt \{rahmanm, bkojusner,ryonkennedy, preritpathak, linqi, byron\}@ufl.edu}\vspace{-1.0ex}}

\maketitle

\begin{abstract}
Frequently advised secure development recommendations often fall short in practice for app developers. Tool-driven (e.g., using static analysis tools) approaches lack context and domain-specific requirements of an app being tested. App developers struggle to find an actionable and prioritized list of vulnerabilities from a laundry list of security warnings reported by static analysis tools. Process-driven (e.g., applying threat modeling methods) approaches require substantial resources (e.g., security testing team, budget) and security expertise, which small to medium-scale app dev teams could barely afford. To help app developers securing their apps, we propose \sorcerer\footnote{Sourcerer is a fictional character depicted in the fantasy novel series `Discworld' written by Terry Pratchett. \url{https://discworld.fandom.com/wiki/Sourcerer}}, a guiding framework for Android app developers for security testing. \sorcerer~ guides developers to identify domain-specific assets of an app, detect and prioritize vulnerabilities, and mitigate those vulnerabilities based on secure development guidelines. We evaluated \sorcerer~ with a case study on analyzing and testing 36 Android mobile money apps. We found that by following activities
guided by \sorcerer, an app developer could get a concise and actionable list of vulnerabilities (24-61\% fewer security warnings produced by \sorcerer~than a standalone static analyzer), directly affecting a mobile money app's critical assets, and devise a mitigation plan. Our findings from this preliminary study indicate a viable approach to Android app security testing without being overwhelmingly complex for app developers.

\end{abstract}

\pagestyle{plain}

\section{Introduction}
\label{intro}
Mobile apps usage is increasing, and security vulnerabilities are corroding apps. In a 2021 mobile security report, Synopsis reported that out of the 3335 top mobile apps analyzed from 18 of the most popular app categories during the Covid-19 pandemic, 63\% of apps contained known security vulnerabilities~\cite{AnalystR84:online}. While apps ridden with security vulnerabilities are nothing new, one interesting observation is that 94\% of these vulnerabilities have publicly-known fixes available, and 73\% of the discovered vulnerabilities were first disclosed more than two years ago. As the onus of securing mobile apps is often placed on app developers, one cannot ignore asking: \textit{why do app developers fail to adopt security tools and guidelines when security practitioners and researchers share their decades of experience dealing with common vulnerabilities?}

The well-known secure development recipes frequently advertised to developers can be broadly categorized as either tool-driven or process-driven. Advocates for tool-driven security recommend that app developers use static analysis tools to detect and fix vulnerabilities extensively. However, developers face challenges in understanding, contextualizing, and prioritizing security warnings from static analysis tools because of high false-positive rates, indecipherable warnings, and context-insensitive reporting~\cite{smith2020, johnson2013don}. On the other hand, while process-driven threat modeling techniques (e.g., STRIDE, Attack Trees, Top-k lists of vulnerabilities) have been proposed and practiced in the industry, various factors like time, budget, dedicated security teams/experts, and a steep learning curve make threat modeling challenging and unadoptable for solo app developers or small teams that do not have structured organizational capacity~\cite{yskout2020threat}. Moreover, no tailored threat modeling process focusing on secure mobile app development is currently available. While app developers are aware of the mobile security knowledge base (e.g., OWASP Mobile Security Project), there is no straightforward recipe for app developers to jump-start the threat modeling process for security testing. Research has shown that solo and small-scale app developers want support to make the `right' choices regarding thinking and implementing security controls~\cite{Linden2020}. We argue that ignoring the need for small scale app developers has already resulted in an app-insecurity problem, including client-side insecure TLS certificate validation (e.g.,~\cite{fahl2012eve}), code injection through insecure Webview implementations (e.g.,~\cite{yang2019iframes}), and privilege escalation attacks using third-party libraries to name a few of the commonly recurring app vulnerabilities.

In an attempt to help app developers contextualize functional
requirements of their apps during security testing, we propose \sorcerer, a lightweight, developer-driven guiding framework for Android app security testing. \sorcerer~ provides an easy-to-follow recipe for app developers to identify critical assets of an app, detect and map prioritized vulnerabilities affecting assets, and mitigate prioritized vulnerabilities based on secure development guidelines. We made the following contributions in this paper:

\vspace{-0.3em}

\begin{enumerate}[ 1.]
	\item We situate the \sorcerer~framework by aligning its threat model and assets under the scope of Android app development~(\S\ref{framework}). We define three phases of activity for an app developer to conduct principled security testing. 
	
	\item We define a taxonomy of asset families for Android app threat modeling, which can be used to identify assets for apps operating in diverse business domains~(\S\ref{android_assets}).
	
	\item We provide a walkthrough of the framework by providing a case study and detailing the walkthrough of applying the framework to test 36 Android mobile money apps~(\S\ref{casestudy}). We present results~(\S\ref{results}) and discuss~(\S\ref{discussion}) potential application areas of \sorcerer~ in app security testing and developer security education.
	
\end{enumerate}

\vspace{-0.5em}

\section{\sorcerer~Framework}
\label{framework}

\sorcerer~is a guiding framework built on the principle of identifying critical assets to prioritize and mitigate vulnerabilities during app security testing. As app development is iterative in nature with shorter release cycles, developers face challenges in testing apps to address devices and OS version compatibility, accessibility, and UX, with little to no consideration for security testing~\cite{ahmad2018empirical, asfour2019exploring}. This gap in security testing motivates us to introduce \sorcerer~ to guide Android app security testing. 

In \S\ref{concept_definition}, we introduce the key concepts of \sorcerer~framework and in \S\ref{sourcerer_phases}, we describe the phases of \sorcerer~ to conduct Android app threat modeling.


\subsection{Definition of Key Concepts}
\label{concept_definition}

\subsubsection{Threat Model for Android App Security}
Recently, Mayrhofer et al.~\cite{mayrhofer2019android} proposed a multi-layered threat model for Android OS, which we adopt to discuss an adversary's capabilities under the scope of developing a secure Android app. We assume that the adversary can sniff network data packets sent/received by an app, access the platform-provided API stack through the app, run code for harvesting sensitive data, and transfer sensitive data to adversary-controlled locations without user notice. The adversary does not gain physical access to an end-user device, but they can install an app in an adversary-controlled device to reverse-engineer the app. We assume the end-user trusts their device and the underlying Android OS. For the scope of this paper, we assume the following threats that an adversary can exploit because of the insecure app development process: (i)~untrusted code execution on the device, (ii)~untrusted content processed and stored by the device, and (iii)~untrusted network communication.

\subsubsection{Assets for Android Devices}
\label{android_assets}

In security threat modeling, an asset refers to any piece of information, device, or component of an organization's systems deemed valuable – often because it holds sensitive data or can be used to access such information~\cite{saitta2005trike,jaatun2008covering}. An asset container refers to the physical or logical location where that information is kept. An adversary seeks to access or impact assets through vulnerabilities. 

Considering assets as sensitive information, Android devices hold a multitude of sensitive data (i.e., assets), which include but are not limited to personally identifiable information (PII), business data or corporate intellectual data, authentication data, device identification data (e.g., IMEI, SIM serial no.), application generated data (e.g., cache data, app logs), usage history data (e.g., SMS/call log, browsing history), sensor data generated by tailored hardware (e.g., camera, GPS, gyroscope), user-generated data (e.g., keystrokes, no. of steps taken by user).

After reviewing threat modeling and Android app analysis literature, we propose the following asset families for Android apps. We assume that members of asset families are not exclusive, and there could be overlap in membership among asset families.

\begin{itemize}
	\item \textbf{User Assets:} refer to any set of information that can be used to uniquely identify a device's user, in the form of identifiers (e.g., name, address, email, social security number (SSN), phone number, date of birth, device identifiers) or quasi-identifiers (e.g., frequently visited locations, spending habit, call logs). Compromising user assets could result in either profiling or tracking users by an adversary.
	
	\item \textbf{Application Assets:} refer to the core building blocks of an app to provide advertised functionalities (e.g., money transfer, fitness tracker) with the assumption of keeping user assets secured. Common examples of application assets include, but are not limited to, cryptographic keys, API keys/tokens to connect web/cloud services, proprietary algorithm/information. Compromised application assets can result in the loss of intellectual property, data breach, and loss of business reputation.
	
	\item \textbf{Platform Assets:} refer to those core functionalities served by the host operating system. For example, in Android, platform assets include, but not limited to, system APIs to perform sensitive operations, e.g., accessing local media storage/files (e.g., \texttt{READ_EXTERNAL_STORAGE}, \texttt{READ_SMS}, \texttt{READ_CALL_LOGS}), accessing rich sensors (e.g., \texttt{CAMERA}, \texttt{ACCESS_FINE_LOCATION}) for customized end-user services.
	
\end{itemize}

\subsection{Phases of \sorcerer}
\label{sourcerer_phases}

We present \sorcerer~as a three-phased guiding framework for security testing of an Android app before the app is published on the marketplace. We assume that an app developer does not require formal security training to follow the process prescribed by the \sorcerer~ framework. We also assume that the developer (or security tester) has complete access to the app source code for code review and business requirements documentation for reviewing app requirements. Followings are the activities involved in the three phases of the \sorcerer~ framework that an app developer needs to follow through for app security testing:

\begin{enumerate}[ 1.]
	\item \textbf{Phase 1 - Asset Identification:} Identification of assets from app requirements documentation.
	\item \textbf{Phase 2 - Vulnerability-to-Asset Mapping}:
	Source code analysis for vulnerability detection and prioritization based on assets.
	\item \textbf{Phase 3 - Mitigation:} 
	Apply mitigations based on the vulnerability-to-asset mapping.
\end{enumerate}

%

\subsubsection{Phase 1 - Identification}
In the first phase, an app developer analyzes the app's functional/business requirements and identifies an Android app's critical assets and asset containers which provide or support certain functionalities in a particular business domain. An asset could belong to multiple asset families based on the app's domain-specific requirements. For example, system-level Android API permissions can belong to both app-asset and platform-asset families, depending on the permission level. Consider a hypothetical fitness app, VeryFiitt, asking a user to grant \texttt{LOCATION} permission to track user's activities. The app needs to safeguard this permission so that no other apps can collude. App collusion happens if VeryFiitt sends the location data via an implicit intent (\textit{action}=\texttt{android.intent.action.SEND}, \textit{MimeType} = \texttt{text/plain}), and a malicious app without location permission can define an intent filter to accept the location intent, leading to escalated privilege violation. In this context, the app developer of the VeryFiitt app should consider \texttt{LOCATION} permission as both an app asset (malicious app collecting user location, thus violating user privacy) and a platform asset (malicious app accessing privileged permission API).

In essence, asset identification can be defined as a function of an event where a system-level/third-party API accesses a sensitive system resource or data type to provide a service defined by the app's business requirements.

%

%
%
%

\subsubsection{Phase 2 - Vulnerability-to-Asset Mapping}

In Phase-2, an app developer focuses on source code analysis to detect vulnerabilities to map the vulnerabilities to assets identified in Phase-1.  For example, a developer uses the source code analysis engine to check whether the app suffers from vulnerabilities such as cross-site scripting (XSS) and remote debugging enabled in \texttt{WebView}. In addition, the source code analysis engine determines whether private data (e.g., user credentials, device identifiers) are logged or sent over an insecure network connection. The source code analysis engine comprises multiple static analysis tools to provide
a comprehensive and comparative security report that examines the codebase from multiple angles. A comparative security analysis is required because static analysis tools differ in sensitivity and vulnerability ranking, and not all static analysis tools can guarantee to cover all types of vulnerabilities~\cite{ranganath2020free}. The comparative security analysis reports vulnerabilities based on \textit{majority-voting} of static analysis tools (a detailed walkthrough of creating a consolidated report based on comparative analysis is described in \S\ref{casestudy}). This list of vulnerabilities is further inspected to verify whether they are exploitable to cause adverse impacts on assets identified on Phase-1. At the end of Phase-2, the developer is able to form a prioritized list of vulnerabilities that could impact assets (i.e., vulnerability-to-asset map).

\subsubsection{Phase 3 - Mitigation}

With the knowledge of vulnerability-to-asset mapping, the app developer can implement security controls and mitigations advocated by bodies of industry-standard groups (e.g., NIST, OWASP, and MITRE). For our discussion, we choose OWASP Mobile Security as our knowledge base to find mitigation techniques for the following reasons: i)~Unlike other standards bodies, the OWASP Mobile Security project primarily focuses on common vulnerabilities that can be exploited client-side (i.e., end-user device). ii)~The vulnerabilities usually arise from implementation-specific errors resulting from developers' misuse, abuse, or misunderstanding of security instrumentation on a mobile device.

\vspace{-0.5em}

\section{Case Study}
\label{casestudy}

To evaluate the \sorcerer framework, we conducted a case study to seek answers to the following questions:

\begin{enumerate}[1.]
	\item RQ1: How long does it take to conduct the three-phase activities of \sorcerer?
	\item RQ2: How does the vulnerability report obtained by following the \sorcerer~ activities differ from a static analysis tool?
	\item RQ3: How do the available open-source static analysis tools compare with each other?
	\item RQ4: What are the common vulnerabilities in mobile money apps? Which assets are getting affected by these vulnerabilities?
	\item RQ5: What are the common mitigations for these vulnerabilities?
\end{enumerate}

Three members from the authors ran the case study by following the prescribed activities of \sorcerer, as defined and explained in~\S~\ref{framework}. All participants had intermediate to advanced level experience in Java programming language and intermediate level experience in Android app reverse engineering.

We chose Android mobile money apps for the case study as these apps have revolutionized people's access to the economy and financial independence in developing countries in the last decade, yet the security posture of these apps is found vulnerable~\cite{reaves2017mo}. Based on GSMA mobile banking report~\cite{gsmaReport2019}, we chose 36 Android fintech apps that are geographically diverse and are currently operational in developing countries from South-East Asia, Middle East, North and Central Africa, Central America, and South America. 

Each participating evaluator was assigned to conduct security testing of twelve apps based on the \sorcerer~ framework. For a comparative static analysis, participants were instructed to use three static analysis tools: MobSF, AndroBugs, and QARK. These tools were chosen because they are open-sourced, well documented, and extensively used in academic and industry app analysis for benchmarking open-source static analysis tools~\cite{reaves2016droid,ranganath2020free}. In the absence of standard project requirement documents, we instructed participants to use the app descriptions available from the Google Play store as a proxy to identify the functional requirements and services provided by the apps. 

We asked each participant to measure time during each phase of the study and to take notes of any challenges faced during the study. \textbf{Since the case study was based on static app analysis, no unexpected damage or any harm was not incurred to the affiliated financial institutions or end-users.} In the following, we describe the activities that a participant, Alice (pseudonym), conducted for security testing of the A2 app (a mobile banking app in India).

\subsection{Phase 1: Asset Identification}
In this stage, Alice manually examined the A2 app's description from the Google Play store to identify the keywords that defined either an action the end-user could perform or the type of service provided by the app. She identified the following text snippets which show the functionalities of the app: 

\textit{$\ldots$instant digital payments through your mobile phone$\ldots$link Mobile number with Bank account$\ldots$phone should have an active SIM card$\ldots$The active SIM card is linked to a bank account$\ldots$You have a valid Debit Card for your Bank Account$\ldots$Use your PIN to unlock device}

Based on these functional use cases, Alice listed the following asset family members:

\begin{itemize}
	\item\textbf{User Assets:} SIM card no., bank account no., PIN, phone no., phone contacts, debit card no.
	\item\textbf{App Assets:} Proprietary access control mechanism (e.g., using PIN) to login the device/app, proprietary cryptographic algorithms to keep data safe, secure communication channel.
	\item\textbf{Platform Assets:} Access to sensitive data via dangerous permission APIs (e.g.,\texttt{READ_CONTACTS}, \texttt{READ_PHONE_STATE}).
\end{itemize}

[Note: Figure~\ref{fig:summary-results} lists the compilation of common assets found across three asset families by our participants.] 

\subsection{Phase 2: Vulnerability-to-Asset Mapping}

Alice used three tools for comparative static analysis: MobSF, AndroBugs, and QARK. Since the security warning reports produced by each tools were heterogeneous in terms of reporting style and vulnerability ranking~\cite{ranganath2020free}, Alice followed this heuristic of \textit{majority-voting} inclusion criteria to reconcile the security reports and create an initial list of vulnerabilities for further inspection: \textit{a vulnerability $V$ would be considered if at least two out of the three tools could detect $V$ in exact source code locations (e.g., same class or methods).}

To verify these vulnerabilities, Alice used a Java decompiler IDE jd-gui\footnote{\url{https://github.com/java-decompiler/jd-gui}} to inspect the source code manually. She checked the reported vulnerable methods and code snippets to verify that they were not part of unreachable code. Alice manually checked the data flow of the vulnerable code location and verified if any assets identified in the previous stage are reachable. For confirming the exploitability of vulnerabilities, she looked up code samples from a vulnerability benchmarking repository named Ghera~\cite{ghera2017}. After verification, she mapped each vulnerability to the asset categories that were found to be affected.

\subsection{Phase 3: Mitigation}

To implement security controls based on the vulnerability-to-asset mapping, Alice browsed and searched by vulnerability keyword (e.g., SSL cert. error, insecure WebView) in OWASP mobile app security verification standard (MASVS)~\cite{masvs2021} guidelines to identify prescribed security requirements. After finding the security requirement, she further searched into OWASP mobile security testing guideline (MSTG)~\cite{mstg2021} to find and implement recommended security practices.

\vspace{-0.5em}

\section{Results}
\label{results}

This section presents the preliminary evaluation results of \sorcerer~ framework, based on the case study.

[\textbf{Note}: For anonymization, we have reported only the aggregated results and avoided publishing the name of any specific mobile banking service. We have reported all our findings, detailed security risk reports, and a detailed guideline of performing security testing based on \sorcerer~ to the related financial organizations, with the expectation that they would enhance the security of their apps.]

\begin{figure}[t]
	\includegraphics[width=0.5\textwidth]{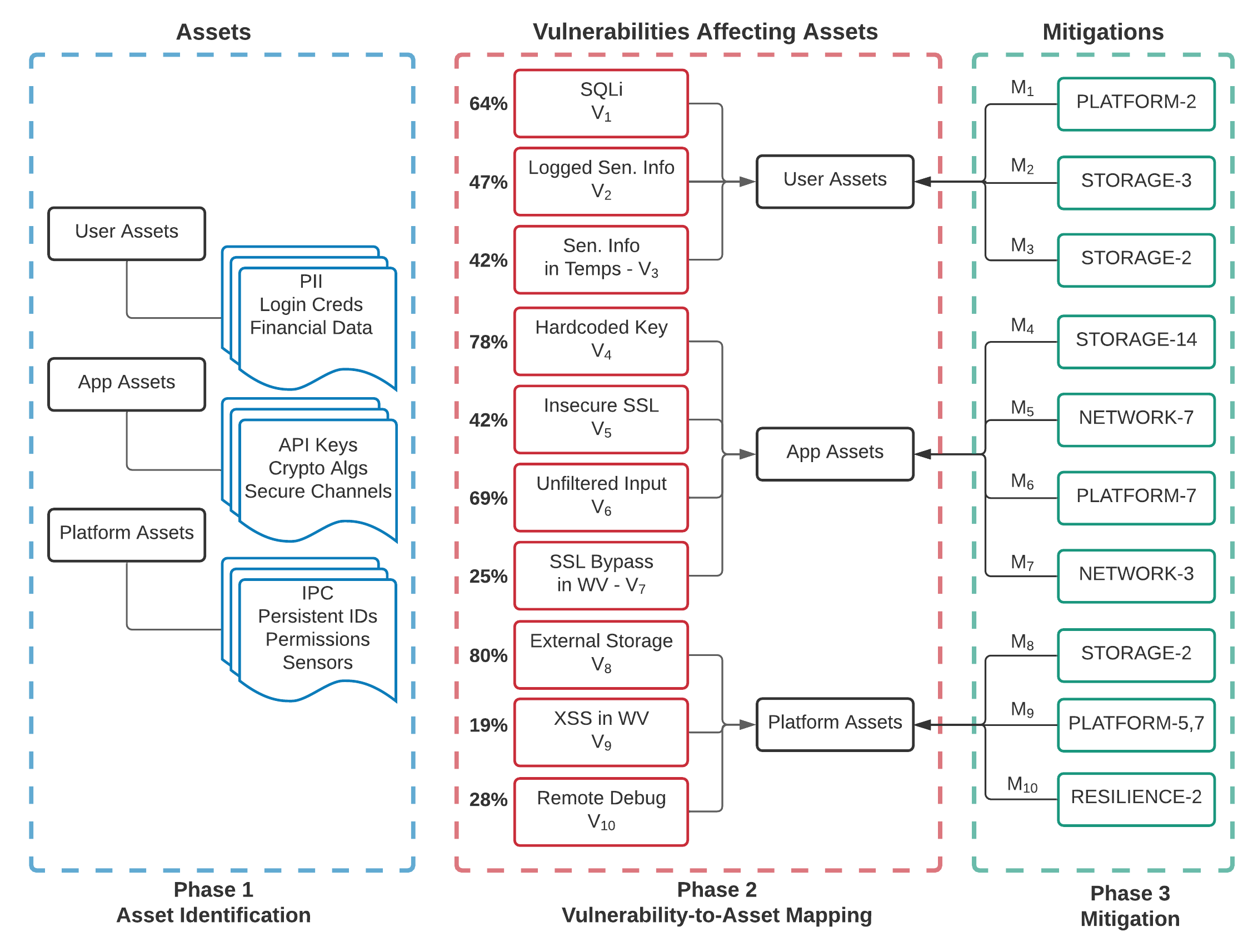}
	\caption{Summary results from the case study. The left column shows the list of assets identified by three participants after performing Phase-1 activities. The Middle column shows the vulnerability-to-asset mapping. The percentages shown next to each vulnerability category show the `\%' of apps affected by the corresponding vulnerability. The rightmost column shows the mitigation strategies as defined in the OWASP MSTG guideline. Mitigation $M_i$ provides implementation strategy to thwart vulnerability $V_j$ (where $i == j$), thus protecting the affected assets. IPC: inter-process communication; PII: personally identifiable information; WV: WebView.}
	\label{fig:summary-results}
\end{figure}

\subsection{RQ1: Runtime for \sorcerer~ Framework}
\vspace{-0.3em}

On average, it took participants \underline{61 minutes} to complete the \sorcerer~ activities, from identifying assets of an app to finding vulnerability mitigations, with a minimum runtime of 51 minutes and a maximum of 75 minutes. 
Participants spent an average of \underline{15 minutes for asset identification}, \underline{30 minutes during vulnerability detection and prioritization}, and \underline{20 minutes to find mitigations}.

\subsection{RQ2: Comparison of \sorcerer~ and Static Analysis Tools}
\vspace{-0.3em}
On average, MobSF, QARK, and AndroBugs reported 9, 19, and 49 categories of vulnerabilities. On the other hand, using the \sorcerer~ framework, our participants could prioritize only seven categories of vulnerabilities on average. Based on participants' reporting, \sorcerer~ effectively reduced prioritized security warnings in a range of 24-61\%, compared to a standalone static analysis tool.

\subsection{RQ3: Comparative Tool Analysis}
\vspace{-0.3em}
Participants reported MobSF as a comprehensive tool for detecting the vulnerabilities found in mobile money apps. The average runtime for MobSF was 50 secs, while the runtime for AndroBugs was 72 secs. QARK was found to have the longest run time of 19 minutes. QARK also failed to analyze 33\% of the apps during the case study. Participants reported both MobSF and AndroBugs being consistent in terms of sensitivity and vulnerability ranking throughout the analysis.

\subsection{RQ4: Common Vulnerabilities and Affected Assets in Mobile Money Apps}
\vspace{-0.3em}
Participants found that 75\% of the apps had known vulnerabilities affecting all three asset categories. Participants found these vulnerabilities reported by static analysis tools, and later they verified the reported vulnerabilities by manual inspection, as reported in \S\ref{casestudy}. SQL injection was the most critical vulnerability affecting user assets (64\% apps), followed by sensitive information (e.g., mobile no., email, IMEI) logging vulnerability (47\% apps). 78\% apps have been found with hardcoded information (e.g., API keys) in plain text, thus affecting app assets. In terms of platform assets, 75\% of the apps have accessed at least one of thirty \textit{dangerous} permissions defined in Android documentation. Interestingly, 63\% of the apps have accessed \texttt{READ_PHONE_STATE} permission, which enables the apps to access \textit{non-resettable unique identifiers}, e.g., IMEI. Accessing identifiers like IMEI is often tied with transmitting sensitive data to an ad-tracking network, which often shares data with data brokers without the explicit consent of end-users~\cite{reardon201950}. Our participants also noticed that 68\% of mobile money apps using third-party tracking libraries (e.g., Tune, Branch, AppsFlyer, Amplitude, Crashlytics, FireBase), as reported by MobSF. While it was beyond the current scope of this paper, this finding certainly raises suspicions as to why sensitive apps like mobile money apps use tracking libraries and perform international data transfer, which may cause violation of federal banking laws of the operating countries~\cite{gsmaReport2019}. 80\% of apps have been found accessing external storage (platform asset). While accessing external storage is not a vulnerability, a close inspection showed that 53\% of apps with external storage access had also logged and stored sensitive information in common storage. The middle column in figure \ref{fig:summary-results} shows summary results of the vulnerabilities identified and the vulnerability-to-asset mapping catalog for 36 analyzed apps.

\subsection{RQ5: Common Mitigations for Mobile Money Apps}
\vspace{-0.3em}
Contrary to the severity of vulnerabilities, mitigations of these vulnerabilities are commonly known and well documented. Most of the vulnerabilities could be mitigated by following secure coding guidelines, e.g., avoiding logging app data in shared storage, avoiding keeping hardcoded data in source code (MSTG-STORAGE-3), verifying that the WebView only rendering JavaScript contained within the app package (MSTG-PLATFORM-7)~\cite{mstg2021}. The rightmost column in figure\ref{fig:summary-results} shows the mitigations identified during evaluation.

\vspace{-0.5em}

\section{Discussion}
\label{discussion}

Mobile app development is an agile process with high-frequency release cycles with a constant rush of adding new features. App developers keep pace with competitors and app market incentives, focusing heavily on fulfilling functional requirements (e.g., device and OS version compatibility) and putting less effort into testing non-functional requirements like security. Commonly available security guidelines are not readily adoptable by Android app developers for the following reasons:
\begin{enumerate}
	\item Unlike legacy systems, android devices pose a diverse set of security and privacy challenges due to their rich set of sensors and always-connected nature over an unsecured network.
	\item The app economy incentivizes developers for faster app publishing in the marketplace and
	\item There is a rapid rise in app developer demographics, from hobbyist to independent, small or medium scale developer teams who cannot afford resources (e.g., time, money, security education) for dedicated security testing.
\end{enumerate}

 Research initiatives have not yet produced any standard, reference model, or exemplar that can be used to capture what threat modeling and security testing entails for secure app development~\cite{yskout2020threat}. Therefore, it is not always clear to app development teams what to expect from app security testing. Despite publishing app security guidelines, OWASP too acknowledges that a one-size-fits-all approach to mobile app security testing is not sufficient because every mobile app is unique and requires a different level of security., e.g., a threat model for a book reader app would be different from that of a money transfer app. We argue that the list of commonly available security requirements and testing guidelines is too broad and can not be readily tailored and adapted by a developer to measure the security posture of her app.  For example, some could believe that addressing the `top 10 threats' is a valid substitute for systematic security testing. However, we believe that a common reference framework will help to clarify the value of systematic security testing for apps.
Furthermore, a standard reference framework serves as the basis for exchanging threat modeling information in a standardized fashion and thus can underpin tools that support the security testing process. Research has shown that small- to medium-scale app developers need tool support and contextualized decision
support regarding thinking and implementing security controls~\cite{Linden2020}. We believe \sorcerer~ can be a guiding framework by bridging this current gap between knowledge and action. \sorcerer~ helps an app developer prioritize and fix vulnerabilities based on the critical domain-specific assets of an app without getting overwhelmed by a generic list of security advice or warnings reported by standards bodies or static analysis tools.

\vspace{-0.5em}

\section{Related Work}

We discuss related work in two key areas: investigations of security
issues in Android app development, and studies of threat modeling 
techniques for secure software development.

There have been some efforts to investigate security vulnerabilities in Android apps by developing static analysis tools. For example, Crylogger~\cite{Luca2021} detects cryptographic API misuses in Android apps; MalloDroid~\cite{fahl2012eve} detects potential SSL Man-In-The-Middle (MITM) vulnerabilities in apps; FixDroid~\cite{nguyen2017stitch} scans a developers’ code for common security pitfalls and provides feedback. It needs
to be pointed out that working prototypes of academic static
analysis tools are hard to find in the real world, thus makes it
challenging and harder for reproducibility and re-evaluation
for future work, let alone to be used for actual app development phases,
as found in systematic studies of Android static analyzers~\cite{reaves2016droid, ranganath2020free}. Besides academic tools, industry-grade static analysis tools also present warnings to developers in a non-user-friendly and unintuitive format, which leads developers to confusion, due to either a lack of security knowledge or person-hours available to address the issues~\cite{johnson2013StaticAnalysis}.

Threat modeling in the context of secure software development is involved in identifying threats to mitigate vulnerabilities. While various
threat modeling approaches (e.g., STRIDE, PASTA) have been proposed and developed over time, the practice is not widespread and agile practitioners, and software developers have few sources available on adopting the threat modeling practice during development~\cite{cruzes2018challenges, yskout2020threat}. For example, asset identification is one of the greatest
challenges faced by development teams~\cite{cruzes2018challenges}; developers need a baseline and a guiding framework to start modeling the system, eliciting security requirements, and finally reviewing the system to apply security controls as required~\cite{yskout2020threat}. Besides, the practice of threat modeling often poses challenges, e.g., steep learning curve, collaboration across teams, and internal or external security expertise. These challenges are critically hard to manage for small (or solo) to medium level app developers with limited to zero levels of organizational resource capacity~\cite{ahmad2018empirical, Linden2020, yskout2020threat}.

In our work, we focus explicitly on the lack of a developer-driven security testing framework and propose \sorcerer~ that would help developers contextualize functional requirements of their apps and prioritize threats for mitigation to secure the apps.

\vspace{-0.5em}

\section{Limitations and Future Work}
\label{limitation}
Due to obfuscated source code, participants from the case study had to use their subjective assessment to choose whether a particular vulnerability would be critical or not. For example, one-third of the apps used weak hash algorithms, but participants could not confirm whether the hash algorithms were part of an encryption mechanism or something benign. Also, the smaller sample size of human evaluators limited our ability to generalize our findings. For example, the average time reported by evaluators to complete \sorcerer~ activities could be biased, as the participants had prior knowledge in reverse engineering, which may not be common to know for a general app developer. In an ideal case, an app developer might take more than an hour to complete the activities. 
Though we believe it is essential, we did not focus on improving vulnerability verification through taint analysis except for leveraging the existing vulnerability benchmark available in the literature (see Phase-2 from \S~\ref{casestudy} ). Relying solely on manual inspection could bias the vulnerability verification results found during the case study. During the reconciliation of security reports from multiple static analysis tools, the participants might disregard valid vulnerabilities due to applying a majority-voting heuristic, which could affect the results to be generalized. Besides, the observed results are based on a limited set of static analysis tools, and hence the observations regarding open-source static analysis tools may fall short in generalization.

In response to these limitations, we point out the fact that our work's primary focus was not to perform \textit{yet another app analysis} to find vulnerabilities. Instead, we proposed a guiding framework that could help app developers have a reference framework for threat modeling and security testing with a minimal learning curve and less friction to their regular development activity. Additionally, although we present a case study of \sorcerer~ on fintech apps, this framework can be readily repurposed to analyze Android apps from any business domain and apps from other smartphones OS (e.g., iOS).

The work-in-progress consists of automating the manual steps (e.g., asset identification, vulnerability verification) and implementing a standard and bespoke tool for app developers to conduct the activities intuitively with less manual effort. Future work will consist of conducting a large-scale study by applying this framework to analyze diverse categories of mobile apps and developer study to measure the effectiveness and usability of \sorcerer~ framework, comparing other off-the-shelf security tools.

\section{Conclusion}
\label{conclusion}
In this paper, we present \sorcerer~ as a guiding framework for Android app developers to conduct security testing. The framework consists of three phases: identifying assets, creating vulnerability-to-assert mapping, and mitigating identified vulnerabilities. We conducted a case study of evaluating \sorcerer~ framework on 36 financial apps with three participants. Our initial findings showed that \sorcerer~ could help app developers identify a concise and actionable list of vulnerabilities that could affect an app's critical assets and devise a mitigation plan. We also found that study participants could complete the activities within a reasonable time frame. Our findings from this preliminary study indicate \sorcerer~ being a viable approach to Android app security testing without being overwhelmingly complex for app developers.

{\footnotesize
	\bibliographystyle{IEEEtran}
	\interlinepenalty=10000 
	\bibliography{references, quals}
}

%

\end{document}